\documentclass{llncs}
\usepackage[dvips]{epsfig}
\usepackage{footmisc}

\title{Compressed String Dictionaries
\thanks{First author partially funded by the Ministry of Science and Innovation of 
Spain (PGE and FEDER) TIN2009-14560-C03-02 and Xunta of Galicia ref. 09TIC060E, third
author by the Ministry of Science and Innovation of 
Spain: TIN2009-14009-C02-02, and last three authors by Millennium 
Institute for Cell Dynamics and Biotechnology (ICDB), Grant ICM P05-001-F, Mideplan, 
Chile. }}

\author{
Nieves R. Brisaboa\inst{1}
\and
Rodrigo C\'anovas\inst{2}
\and
Miguel A. Mart\'{\i}nez-Prieto\inst{2,3}
\and
Gonzalo Navarro\inst{2}
}

\institute{
Database Lab, Universidade da Coru\~na, Spain
\and
Department of Computer Science, University of Chile
\and
Department of Computer Science, Universidad de Valladolid, Spain
}

\begin{document}
\maketitle

\begin{abstract}
The problem of storing a set of strings --- a {\em string dictionary} ---
in compact form appears naturally in many cases. While classically it has
represented a small part of the whole data to be processed (e.g., for
Natural Language processing or for indexing text collections), more recent
applications in Web engines, Web mining, RDF graphs, Internet routing, 
Bioinformatics, and many others, make use of very large string dictionaries,
whose size is a significant fraction of the whole data. Thus novel approaches
to compress them efficiently are necessary. In this paper we experimentally 
compare time and space performance of some existing alternatives, as well as 
new ones we propose. We show that space reductions of up to 20\% of the original
size of the strings is possible while supporting fast dictionary searches.
\end{abstract}

\section{Introduction}

String dictionaries arise naturally in a large number of applications. We
associate them classically to Natural Language (NL) processing: finding the 
{\em lexicon} of a text corpus is the first step in analyzing it 
\cite{MS99}. They also arise, together with {\em inverted 
indexes}, when indexing text collections formed by NL \cite{BYRN99,WMB99}.

In those NL applications, there has not been much concern about the size of 
the dictionary. This is because, in classical NL collections, the dictionary 
grows sublinearly with the text size: Heaps' law \cite{Hea78} establishes that 
in a text of length $n$, the dictionary size is $O(n^\beta)$, for some
$0<\beta<1$ depending on the type of text. This $\beta$ value is usually 
in the range 0.4--0.6 \cite{BYRN99}, and thus the dictionary of terabyte-size
collections should occupy just a few megabytes and would easily fit in the 
main memory of a commodity PC.

Heaps' law, however, does not model well the reality of Web
search engines. Web collections are much less ``clean'' than text collections
whose content quality is carefully controlled. Dictionaries of Web crawls 
easily exceed the gigabytes, due to typos and unique identifiers that are taken
as ``words'', but also for ``regular words'' from multiple languages. The  
\textit{ClueWeb09} 
dataset\footnote{\tt http://boston.lti.cs.cmu.edu/Data/clueweb09} is a real
example which comprises close to 200 million different words obtained from 1 
billion web pages on 10 languages. This results in a large dictionary 
of far more than 1GB.

Web graphs are another application where the size of the URL names, 
classically neglected, is becoming very relevant with the advances of the
techniques that compress the graph topology. The nodes of a Web graph are
typically the pages of a crawl, and the edges are the hyperlinks.
Typically there are 15 to 30 links per page. Compressing Web
graphs has been an area of intense study, as it permits caching larger graphs
in main memory, for tasks like Web mining, Web spam detection, finding
communities of interest, etc. \cite{KKRRT99,donato2006algorithms}. 
In several cases the URL names
are used to improve the mining quality \cite{YGLS05,Nag10}.

In an uncompressed graph, 15 to 30 links per page would require 60 to 120
bytes if represented as a 4-byte integer. This posed a more serious memory
problem compared to the name of the URL itself once some simple compression
procedure was applied to those names (such as Front-Coding, see 
Section~\ref{sec:frontcoding}). For example,
Broder et al. \cite{BKMRRSTW00} reports 27.2 bits per edge (bpe) and 80 bits 
per node (bpn). This means that each node takes around 400--800 bits to
represent its links, compared to just 80 bits used for storing its URL.
In the same way, an \textit{Internet Archive} graph of 115M nodes and
1.47 billion edges required \cite{SY01} 13.92 bpe plus around 50 
bpn, so 200--400 bits are used to encode the links and only 50 for the
URL. In both cases, the space required to encode the URLs was just 10\%-25\%
of that required to encode the links. However, the advances in compressing 
the edges has been impressive in recent years, achieving compression ratios 
around 1--2 bits per edge \cite{BV04,AD09}. At this rate, the edges 
leaving a node require on average 2 to 8 bytes, compared to which the name of 
the URL certainly becomes an important part of the overall space.

Another application is Bioinformatics. Popular alignment software like
BLAST \cite{G07} indexes all the different substrings of length
$q$ of a text, storing the positions where they occur in the sequence database.
For DNA sequences $q=11,12$ is common, whereas for proteins they use $q=3,4$.
Over a DNA alphabet of size 4, or a protein alphabet of size 20, this amounts
to up to 200 million characters. Using a larger $q$ would certainly 
allow them improve the quality in searching for conserved regions, but this
is infeasible for memory constraints.

The emergent \textit{Linked Data Project}\footnote{\tt http://linkeddata.org}
focuses on the publication of 
RDF\footnote{\tt http://www.w3.org/TR/rdf-syntax-grammar}
data and their connection between different data sources in the ``Web of 
Data''. This movement results in huge and heterogeneous RDF 
datasets from diverse fields.

The dictionary is an essential component in the logical division of an RDF 
database \cite{FMPG10}. However, its effective representation has not been
studied in depth. 
Our experience with the tool 
\texttt{HDT-It!}\footnote{\tt http://code.google.com/p/hdt-it} shows that
the dictionary for the DBpedia-en 
dataset\footnote{\label{fn:dbpedia} \tt http://downloads.dbpedia.org/3.5.1/en}%
takes around
5.14GB, whereas the compact ``\textit{BitmapTriples}'' representation of 
their triples structure only takes 1.08GB. That is, the dictionary amounts 
to more than 80\% of the total structure size.

Finally, Internet routing poses another interesting problem on dictionary 
strings. Domain name servers map domain names to IP addresses, and routers map
IP addresses to physical addresses. They may handle large dictionaries of
domain names or IP addresses, and serve as many request per second as possible. 

This short tour over various example applications shows that handling
very large string dictionaries is an important and pervasive problem. 
Curiously, we have not seen much research on compressing
them, perhaps because a few years ago the space of these dictionaries was
not a serious problem, and at most Front-Coding was sufficient. In this paper
we study Front-Coding and other solutions we propose for compressing large
string dictionaries, so that two basic operations are supported: (1) given 
a string, give its position in the dictionary or tell it is not in the
dictionary; (2) given a position, retrieve its string content. 

Our study over various application scenarios spots a number of known 
and novel alternatives that dominate different niches of the 
space/time tradeoff map. The least space-consuming variants perform efficiently 
while compressing the dictionary to 12\%--30\% of its original size, depending
on the type of dictionary.

\section{Basic Concepts and Related Work}

\subsection{Rank and Select on Bitmaps}
Let $B[1,n]$ be a $0,1$ string ({\em bitmap}) of length $n$ and assume there are $m$ 
ones in the sequence. We define $rank_b(B,i)$ as the number of occurrences of bit $b$
in $B[1,i]$ and $select_b(B,i)$ as the position of the $i$-$th$ occurrence of 
$b$ in $B$.

In this paper we will use two different succinct data structures%
\footnote{Implementations available at {\tt http://libcds.recoded.cl}} 
that answer 
{\em rank} and {\em select} queries. The first one, that we will refer to as 
{\em RG} \cite{GGMN05}, uses $(1+x)n$ bits to represent $B$. It supports {\em 
rank} using two random accesses to memory plus $4/x$ contiguous (i.e., cached)
accesses. An additional binary search is needed to support $select$.

The second data structure, that we will call {\em RRR} \cite{RRR02}, is a 
compressed bitmap that uses in practice about $\log {n \choose m} + 
(\frac{4}{15}+x)n$ bits\footnote{Our logarithms are in base 2.}, 
answering {\em rank} within two random accesses plus
$3+8/x$ accesses to contiguous memory, and {\em select} with an extra binary
search. In practice this compresses the bitmap when $m < 0.2\,n$.

\subsection{Huffman and Hu-Tucker Codes}
\label{sec:coding}

For compressing sequences, statistical methods assign
shorter codes (i.e., bit streams) to more frequent symbols. Huffman coding 
\cite{Huf52} is the optimal code (i.e., it achieves the minimum length of 
encoded data) that is uniquely decodable. In this paper we use canonical 
Huffman codes \cite{MK95}, which have various advantages.

Hu-Tucker codes \cite{K73} are 
optimum among those that maintain the lexicographical order of the symbols. 
Two sequences encoded using Hu-Tucker can be lexicographically compared
bytewise directly in encoded form. We use both codes in this paper, 
in some cases padding them (with zeros) to the next byte in order to simplify 
alignment and bytewise comparisons.

\subsection{Hashing}
\label{sec:hash}

Hashing \cite{CLRS01} is a folklore method to store a dictionary of any kind.
A {\em hash function} transforms the elements into indexes in a {\em hash
table}, where the corresponding value is to be inserted or sought. A
{\em collision} arises when two different elements are mapped to the same 
array cell.
In this paper we use {\em closed hashing}: If the cell where an element is to 
be found is occupied, one successively probes other cells until finding a free 
cell (insertions and unsuccessful searches) or until finding the element
(successful searches). 

We will consider two policies to determine the next cells to probe when a
collision is detected at cell $x$. {\em Double hashing} computes another 
hash function $y$ that depends on the key and probes $x+y$, $x+2y$, etc. modulo
the table size. {\em Linear probing} is a simpler policy. It tries the 
successive cells of the hash table, $x+1$, $x+2$, etc. modulo the table size. 

The {\em load factor} is the fraction of occupied cells, and it influences 
space usage and time performance. Using good hash functions, insertions and 
unsuccessful searches require on average $\frac{1}{1-\alpha}$ probes with
double hashing, whereas successful searches require 
$\frac{1}{\alpha}\ln\frac{1}{1-\alpha}$ probes. Linear probing requires more
probes on average: $\frac{1}{2}\left(1+\frac{1}{(1-\alpha)^2}\right)$ for 
insertions and unsuccessful searches, and 
$\frac{1}{2}\left(1+\frac{1}{1-\alpha}\right)$
for successful searches. Despite its poorer complexities, we consider also
linear probing because it has advantages on some compressed
representations we try.

\subsection{Front-coding}
\label{sec:frontcoding}

Front-coding \cite{WMB99} is the folklore compression technique for 
lexicographically sorted dictionaries. It is based on the fact that 
consecutive entries are likely to share a common prefix. Each entry
in the dictionary is be differentially encoded with respect to the
preceding one. Two values are used: an integer which encodes the length of 
their common prefix, and the remaining suffix of the current entry.

To allow searches, Front-Coding partitions the dictionary into buckets,
where the first element is explicitly stored and the rest are differentially 
encoded. This allows the dictionary to be efficiently searched using a two-step
process: first, a binary search on the first entry of the buckets locates the
candidate bucket, and second a sequential scan of this candidate bucket rebuilds
each element on the fly and compares it with the query. The bucket size 
yields a time/space tradeoff.

Front-coding has been sucessfully used in many applications. We
emphasize its use in WebGraph\footnote{\tt http://webgraph.dsi.unimi.it}
to encode URL dictionaries from Web graphs.

\subsection{Compressed Text Self-Indexes}

A compressed text {\em self-index} takes advantage of the compressibility of a 
text $T[1,N]$ in order to represent it in space close to that of the 
compressed text, while supporting random access and search operations. More
precisely, a self-index supports at least operations 
$extract(i,j)$, which returns $T[i,j]$, and $locate(p)$, which returns 
the positions in $T$ where pattern $p$ occurs.

There are several self-indexes \cite{NM07,FGNV08}. For this paper we are 
interested in particular in the {\em FM-index} family \cite{FM00,FMMN07}, which
is based on the {\em Burrows-Wheeler transform (BWT)} \cite{BW94}. FM-indexes 
achieve the best compression among self-indexes and are very fast to determine 
whether $p$ occurs in $T$. Many self-indexes are implemented in the {\em
PizzaChili} site\footnote{{\tt http://pizzachili.dcc.uchile.cl}}.

The BWT of $T[1,N]$, $T^{bwt}[1,N]$, is a permutation of its symbols. If the
{\em suffixes} $T[i,N]$ of $T$ are sorted lexicographically, then $T^{bwt}[j]$ 
is the character preceding the $j$th smallest suffix. We use the BWT properties
in this paper to represent a dictionary as the FM-index of a text $T$.

FM-indexes support two basic operations on $T^{bwt}$.
One is the {\em LF-step}, which moves from $T^{bwt}[j]$
that corresponds to the suffix $T[i,N]$ to $T^{bwt}[j']$ that corresponds
to the suffix $T[i-1,N]$ (or $T[N,N]$ if $i=1$), that is $j'=LF(j)$. The
second is the {\em backward step}, which moves from the lexicographical 
interval $T^{bwt}[sp,ep]$ of all the suffixes of $T$ that start with string
$x$ to the interval $T^{bwt}[sp',ep']$ of all the suffixes that start with
$cx$, for a character $c$, that is, $(sp',ep') = BWS(sp,ep,c)$.

\subsection{Grammar-Based Compression}
\label{sec:grammar}
Grammar-based compresson is about finding a small grammar that generates a
given text \cite{CLLPSS05}. These methods exploit repetitions in the text to
derive good grammar rules, so they are particularly suitable for texts
containing many identical substrings. Finding the smallest grammar for a given 
text is NP-hard \cite{CLLPSS05}, so grammar-based compressors look
for good heuristics. We use Re-Pair \cite{LM00} as a concrete compressor, as
it runs in linear time and yields good results in practice.

Re-Pair finds the most-repeated pair $xy$ in the text and replaces all its 
ocurrences by a new symbol $R$. This adds a new rule $R \rightarrow xy$ to the 
grammar. The process iterates until all remaining pairs are unique in the text.
Then Re-Pair outputs
the set of $r$ rules and the compressed text, $\mathcal{C}$. We use a public 
implementation%
\footnote{\texttt{http://www.dcc.uchile.cl/gnavarro/software}} for the 
compressor,
and store rules as a pair of integers taking $\log (\sigma+r)$ bits each, and
symbols of $\mathcal{C}$ using also $\log(\sigma+r)$ bits.

\subsection{Variable-Length and Direct-Access Codes}
\label{sec:dac}

Brisaboa et al.~\cite{BLN09} introduce a symbol reordering technique called 
directly addressable variable-length codes ({\em DACs}). Given a concatenated
sequence of variable-length codes, DACs reorder the target symbols so that
direct access to any code is possible. The overhead is at most one bit per
target symbol, which is not too much if the target alphabet is large.

All the first symbols of the codes are concatenated in a first array $A_1$.
A bitmap $B_1$ stores one bit per code in $A_1$, marking with a 1 the codes
of length more than 1. The second symbols of the codes of length more than
one are concatenated in a second array $A_2$, with $B_2$ marking which are
longer than two, and so on.
To extract the $i$th code, one finds its first symbol in $A_1[i]$. If 
$B_1[i]=0$, we are done. Otherwise we continue in $A_2[rank_1(B_1,i)]$, and
so on.

A variable-length coding we use in this paper (albeit not in combination with
DACs) is Vbyte \cite{WZ99}. It is used to represent numbers of distinct 
magnitudes, where most are small. Vbyte partitions the bits into 7-bit chunks 
and reserves the last bit of each byte to signal whether the number continues 
or not. 

\section{Compressed Dictionary Representations}

We describe now various approches for representing a dictionary within 
compressed space while solving two operations on it:
\begin{description}
\item \textbf{{\em locate($p$)}}: gives a unique nonnegative identifier for 
the string $p$, if it appears in the dictionary; otherwise it returns $-1$.
\item\textbf{{\em extract($i$)}}: returns the string with identifier $i$ in 
the dictionary, if it exists; otherwise returns $NULL$.
\end{description}

\subsection{Hashing and Compression}

We explore several combinations of hashing and compression. We Huffman-encode
each string and the codes are concatenated in byte-aligned form. We insert the 
(byte-)offsets of the encoded strings in a hash table.
The hash function operates over the encoded strings (seen as a sequence 
of bytes, that is, we compare them bytewise). This lowers the time to compute 
the function and to compare search keys (as the string is shorter).
For searching we first Huffman-encode the search string and pad it 
bits to an integral number of bytes.

Our main hash function is a modified Bernstein's hash%
\footnote{{\tt http://www.burtleburtle.net/bob/hash/doobs.html}.
We initialize $h$ as a large prime and replace the 33 by $2^{15}+1$, taking
modulo the table size at each iteration.}. The second function for double 
hashing is the ``rotating hash'' proposed by Knuth \cite[Sec. 6.4]{K73}%
\footnote{Precisely, the variant at 
{\tt http://burtleburtle.net/bob/hash/examhash.html}. We also
initialize $h$ as a large prime.}.

We concatenate the strings in the same order they are finally stored in the 
hash table. This improves locality of reference for linear probing, and gives
other benefits, as seen later (in particular we easily know the length in
bytes of each encoded string). We consider three variants to 
represent the hash table, and combine each of them with linear probing 
{\em (lp)} or double hashing {\em (dh)}. 

The first variant, {\em Hash}, 
stores the hash table in classical form, as an array $H[1,m]$ pointing to the
byte offset of the encoded strings. To answer {\em locate($p$)} we proceed as
usual, returning the offset of $H$ where the answer was found, or $-1$ if not. 
To answer {\em extract($i$)}, we simply decompress the string pointed from 
$H[i]$. Then with load factor $\alpha=n/m$ ($n$ being the number of strings in 
the dictionary), the structure requires $m$ integers in 
addition to the Huffman-compressed strings.

The second variant, {\em HashB}, stores $H[1,m]$ in compact form, that
is, removing the empty cells, in an array $M[1,n]$.
It also stores an {\em RG}-encoded bitmap $B[1,m]$
that marks with a 1 the nonempty cells of $H$. Then $H[i]$ is empty if 
$B[i]=0$, and if it is nonempty then its value is $H[i]=M[rank_1(B,i)]$. Now
{\em locate($p$)} returns positions in $M$, so our identifiers become 
contiguous in the range $[1,n]$, which is desirable. For {\em extract($i$)}
we simply decompress the string pointed from $M[i]$.
The space of this representation is $n$ integers plus $(1+x)m$ bits, where
$x$ is the parameter of bitmap representation {\em RG}. The $n$ integers
require $n\log N$ bits, where $N$ is the total byte length of the encoded 
strings.

The price is in time, as each new probe requires an additional $rank$ on $B$.
However, with linear probing, $rank$ needs to be computed only once, as the
successive cells are also successive in $M$. We only need to access the bits
of $B$ to determine where is the next empty cell.

The third variant, {\em HashBB}, also stores $M$ and $B$ instead of $H$,
but $M$ is replaced by a second bitmap. Note that since we have reordered the
codes according to where they appear in $H$ (or $M$), the values in these 
arrays are increasing. Thus instead of $M$ we store a second bitmap $Y[1,N]$,
where a 1 marks the beginning of the codes. Then $M[i] = select_1(Y,i)$. Bitmap
$Y$ is encoded in compressed form ({\em RRR}). Now the $n\log N$ bits of $M$
are reduced to $\log {N \choose n} + (\frac{4}{15}+x)N$ bits, which is smaller
unless the encoded strings are long.

The price is, again, in time. Each access to $M$ requires a $select$ operation.
Note that linear probing does not save us from successive $select$ operations,
despite the involved string being contiguous, because we have no way to know
where a code ends and the next starts.
\subsection{Front-Coding and Compression}

We consider two variants of Front-Coding.
\textit{Plain Front-Coding} implements the original technique by using Vbyte 
to encode the length of the common prefix. The remaining suffix is 
terminated with a zero-byte. Only bytewise operations are needed to search.
The block sizes are measured in number of strings, so {\em extract($i$)} 
determines the appropriate block with a simple division, and then scans the
block to find the corresponding string.

\textit{Hu-Tucker Front-Coding} is similar, but all the strings and Vbyte 
codes are encoded together using a single Hu-Tucker code. The bucket starts
with the Hu-Tucker code of the first string, which is padded to the next byte 
boundary and preceded by the byte length of the encoded string, in Vbyte form.
This prelude enables binary searching the first strings without decompressing 
them. The rest of the bucket is Hu-Tucker-compressed and bit-aligned, and is 
sequentially decompressed when scanning the bucket, both for locating and for 
extracting. We use a pointer-based Hu-Tucker tree implementation.

\subsection{FM-Index Based Representation}

We use two variants of the FM-index. The first is the {\em SSA} 
\cite{FMMN07}, which compresses $T$ to its zero-order entropy, more 
precisely to $N(H_0(T) + 1)(1 + o(1))$ bits. A second variant, which we call
{\em SSA$^*$}, achieves ``implicit compression boosting'' \cite{MN07}
and reaches higher-order compression, more precisely 
$NH_k(T) + o(N\log\sigma)$ for any $k \le \alpha \log_\sigma N$ and constant
$0<\alpha<1$, where $\sigma$ is the alphabet size of $T$ and $H_k$ is the
$k$-th order empirical entropy \cite{Man01}. The SSA is at {\em PizzaChili} 
and SSA$^*$ is obtained by changing {\em RG} by {\em RRR} in its bitmaps.
Both FM-index implementations support functions $LF$ and $BWS$, as well as
obtaining $T^{bwt}[j]$ given $j$, in time $O(\log\sigma)$. We use the indexes
with no extra sampling because we need only limited functionality.

We concatenate all the strings in lexicographic order, terminating each one 
with a special character, \$, that is lexicographically smaller than all the 
symbols in $T$ (in practice \$ is the ASCII code zero, which is the natural
string terminator). We also add \$ at the beginning of the sequence. Thus we
can speak of the $i$th string in lexicographical or positional order,
indistinctly.

Note that, when the suffixes of $T$ are sorted lexicographically, the first 
corresponds to the final \$, and the next $n$ correspond to the \$s that
precede each dictionary string. Thus $T^{bwt}[1]$ is the final character
of the $n$th dictionary string, and $T^{bwt}[i+2]$ is the final character
of the $i$th string, for $1 \le i < n$. Therefore $extract(i)$ can be
carried out by starting at the corresponding position of $T^{bwt}$ and
using LF-steps until reaching a \$. The $T^{bwt}[j]$ characters traversed spell
out the desired dictionary string in reverse order.

To answer {\em locate($p$)} we just need to determine whether $\$p\$$ occurs 
in $T$. Thus we start with $(sp,ep) = (1,n+1)$ and use $|p|+1$ backward steps 
until finding the lexicographical interval $(sp',ep')$ of the suffixes that 
start with $\$p\$$. If $p$ exists in the dictionary and is the $i$th string, 
then $sp'=ep'=i+1$ and we simply return $i$; otherwise $sp'>ep'$ holds at
some point in the process.

\subsection{Re-Pair Based Representation}

We concatenate all the dictionary strings in lexicographic order and apply 
Re-Pair compression to the concatenation. However, we avoid forming rules that
contain the string terminator. This ensures that each string is encoded with 
an integral number of symbols in $\mathcal{C}$ and thus decompression is fast.

Locating is done via binary search, where each dictionary string to compare
must be decompressed first. We decompress the string only up to the point
where the lexicographical comparison can be decided. For extraction we simply 
decompress the desired string. 

For both operations we need direct access to the first symbol of the $i$th 
string in $\mathcal{C}$. Each compressed string can be seen as a 
variable-length sequence of symbols in $\mathcal{C}$, where they are
concatenated. Thus we use the DAC representation on those sequences. This 
gives fast direct access to the $i$th string, at the price of 1.25 bits per 
symbol: we use {\em RG} representation with 25\% overhead.

\section{Experimental Results}

We consider four dictionaries that are representative of
relevant applications:

\begin{description}
\item{\bf Words} comprises all the different words with at least 3 ocurrences
in the \texttt{ClueWeb09} dataset\footnote{{\tt
http://boston.lti.cs.cmu.edu/Data/clueweb09}; thanks to Leonid Boystov.}.
It contains 25,609,784 words and occupies 256.36 MB.
\item{\bf DNA} stores all subsequences of 12 nucleotides found in the 
sequences of S. Paradoxus published in the \texttt{para} dataset%
\footnote{\tt http://www.sanger.ac.uk/Teams/Team71/durbin/sgrp}.
It contains 9,202,863 subsequences and occupies 114.09 MB.
\item{\bf URLs} corresponds to a 2002 crawl of the {\tt .uk} domain from the
{\tt WebGraph} framework\footnote{\tt http://law.dsi.unimi.it/webdata/uk-2002}.
It contains 18,520,486 URLs and occupies 1.34 GB.
\item{\bf URIs} contains all different URIs used in the 
\texttt{DBpedia-en} RDF dataset%
\footref{fn:dbpedia}
(blank nodes and literals excluded). It contains 30,176,012 URIs and takes
1.52 GB.
\end{description}

We use an Intel Core2 Duo processor at 3.16 GHz, 
with 8 GB of main memory and 6 MB of cache, running Linux kernel 2:6:24-28.
We ran {\em locate} experiments for successful and unsuccessful searches.
For the former we chose 10,000 dictionary strings at random. For the latter we
chose other 1,000 strings at random and excluded them from the indexing. For 
{\em extract} we queried 10,000 random numbers between 1 and $n$. Each data 
point is the average user time over 10 repetitions.

\begin{figure}[!ht]
  \centering
  \includegraphics*[bb=51 47 405 298, scale=0.48]{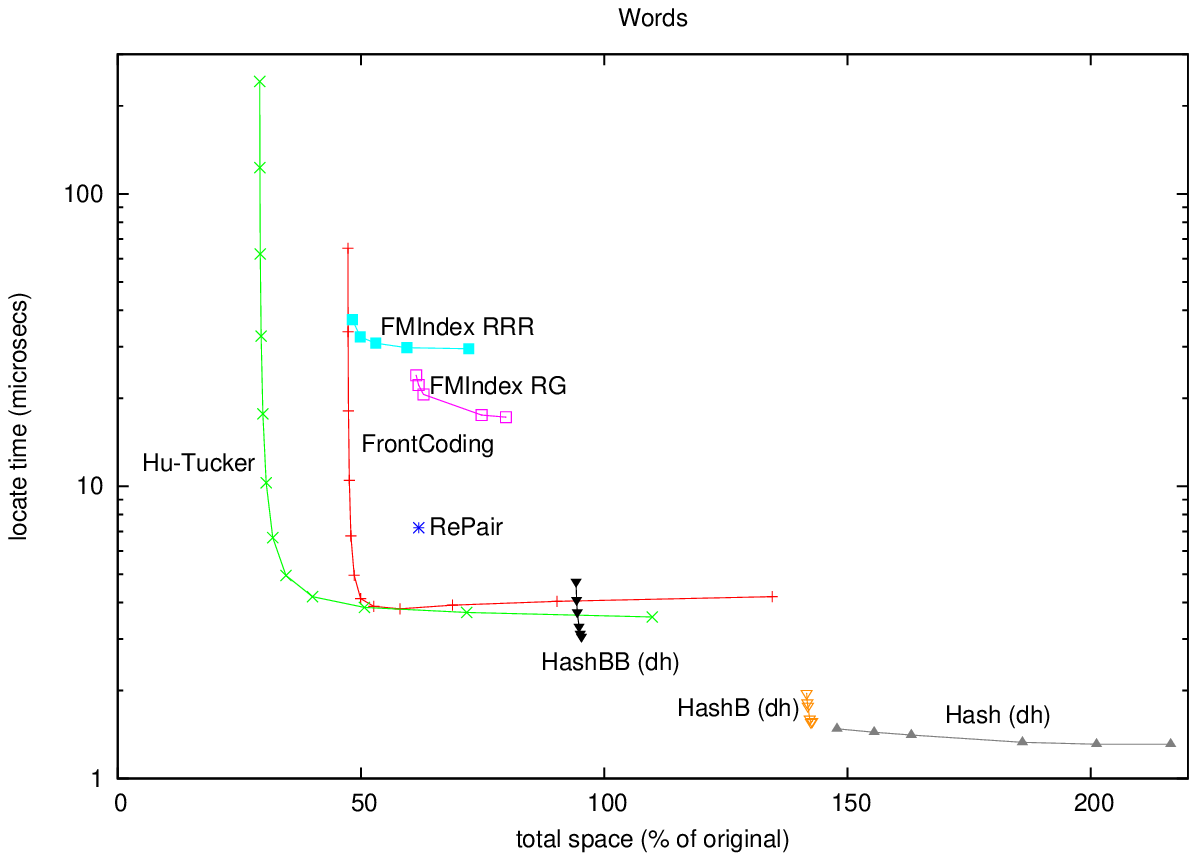}
  \includegraphics*[bb=51 47 405 298, scale=0.48]{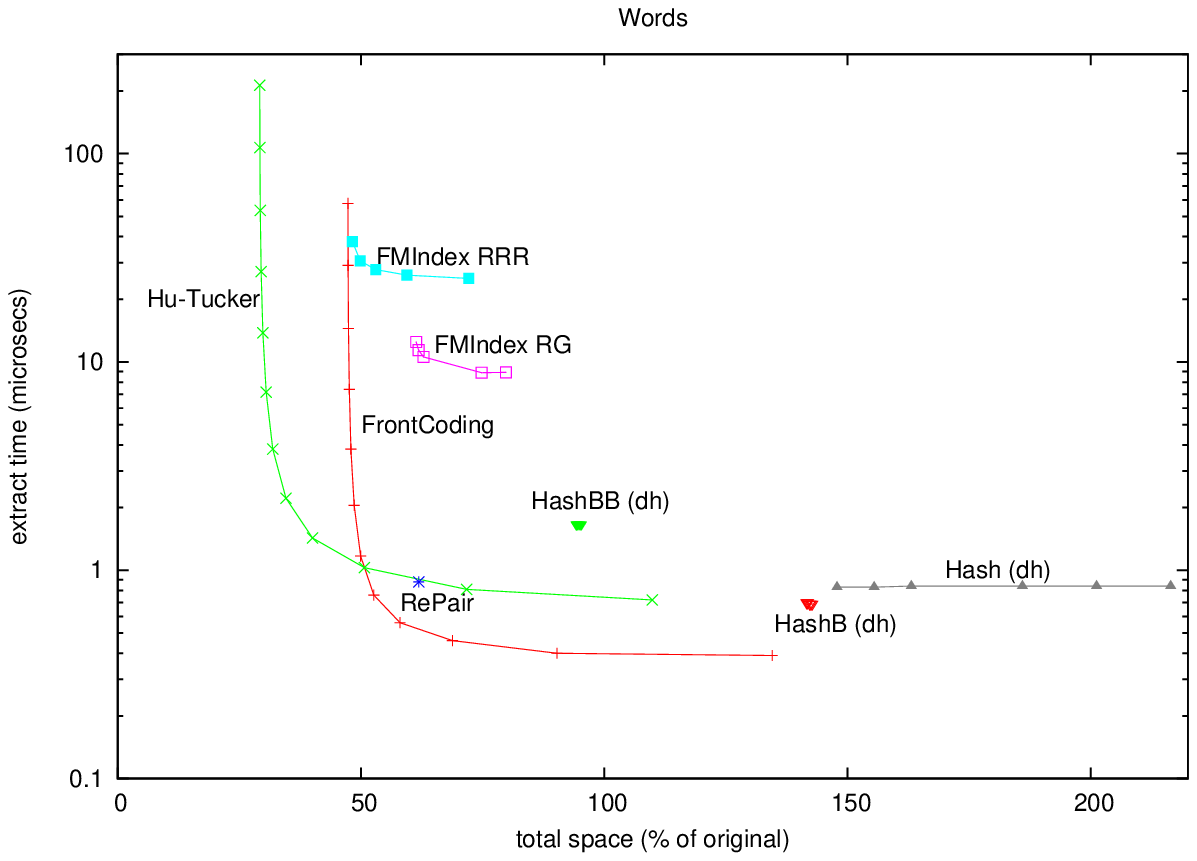}
  \includegraphics*[bb=51 47 405 298, scale=0.48]{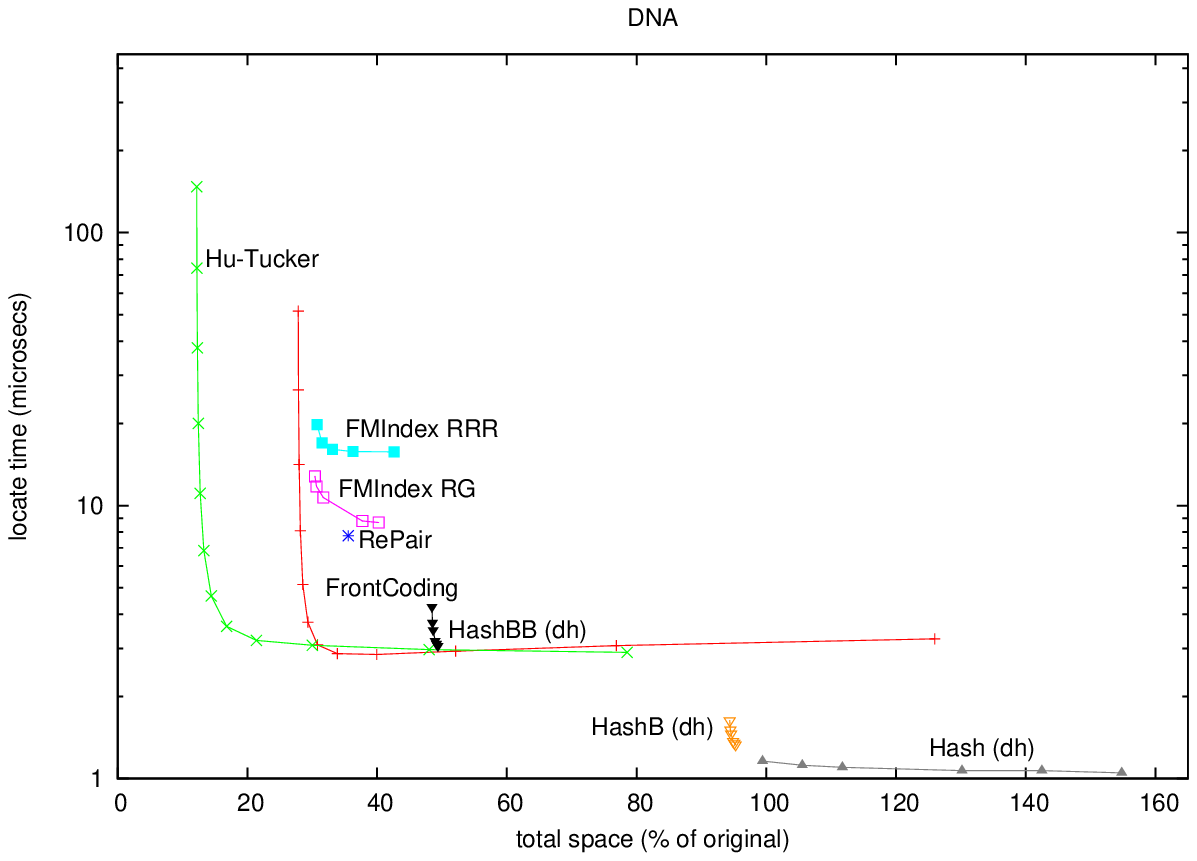}
  \includegraphics*[bb=51 47 405 298, scale=0.48]{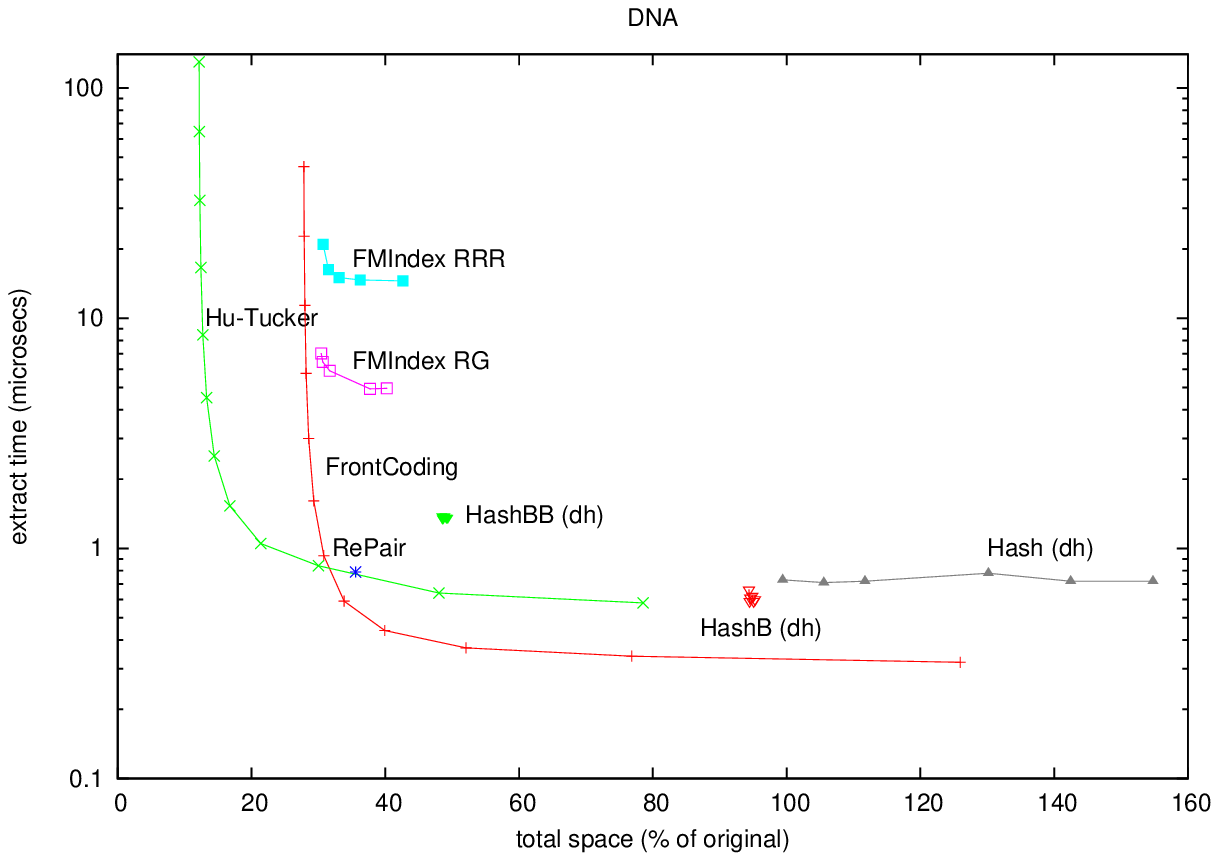}
  \includegraphics*[bb=51 47 405 298, scale=0.48]{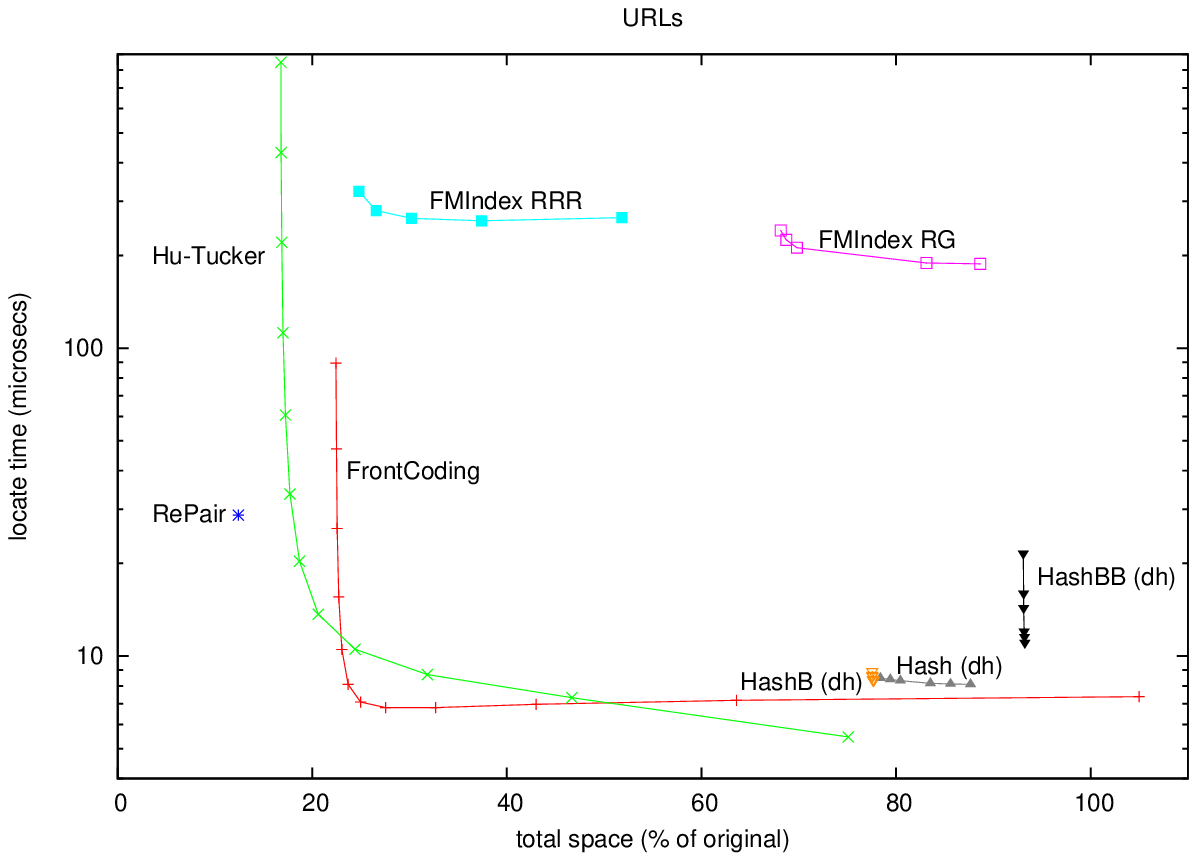}
  \includegraphics*[bb=51 47 405 298, scale=0.48]{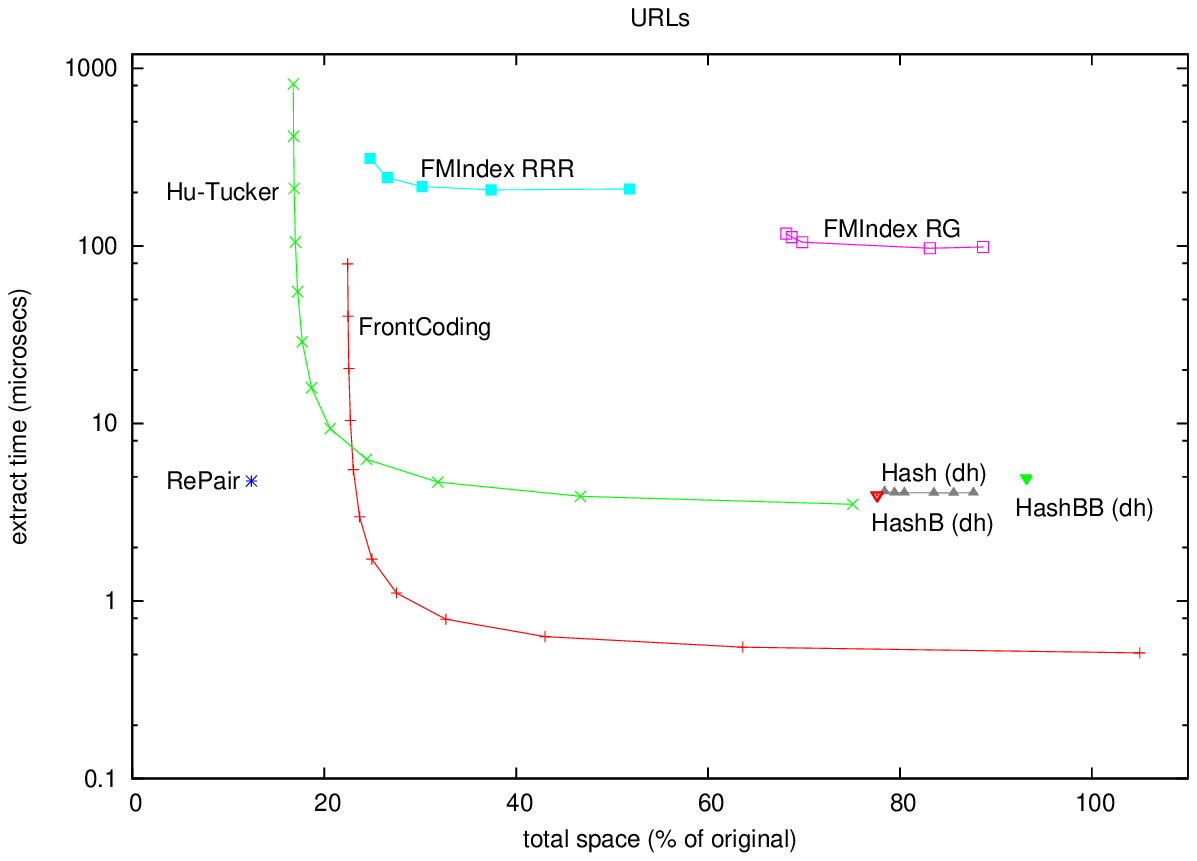}
  \includegraphics*[bb=51 47 405 298, scale=0.48]{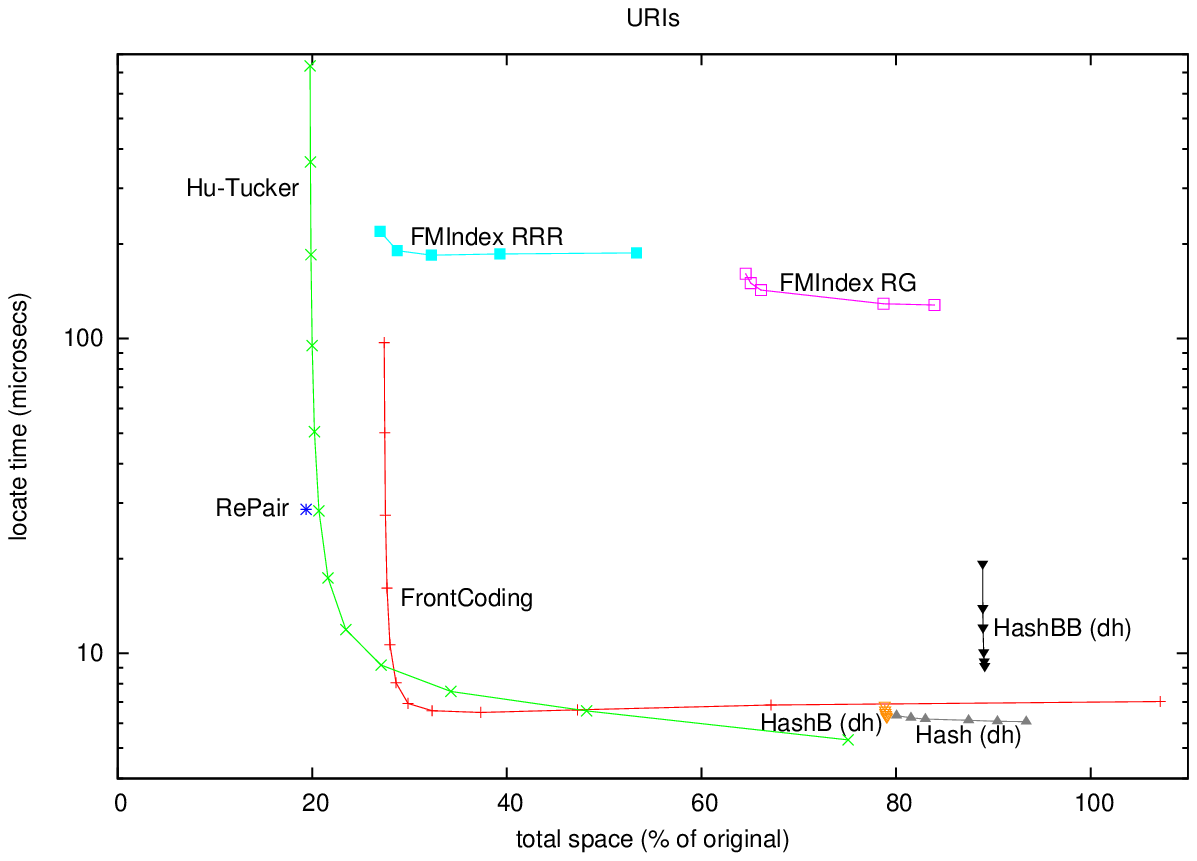}
  \includegraphics*[bb=51 47 405 298, scale=0.48]{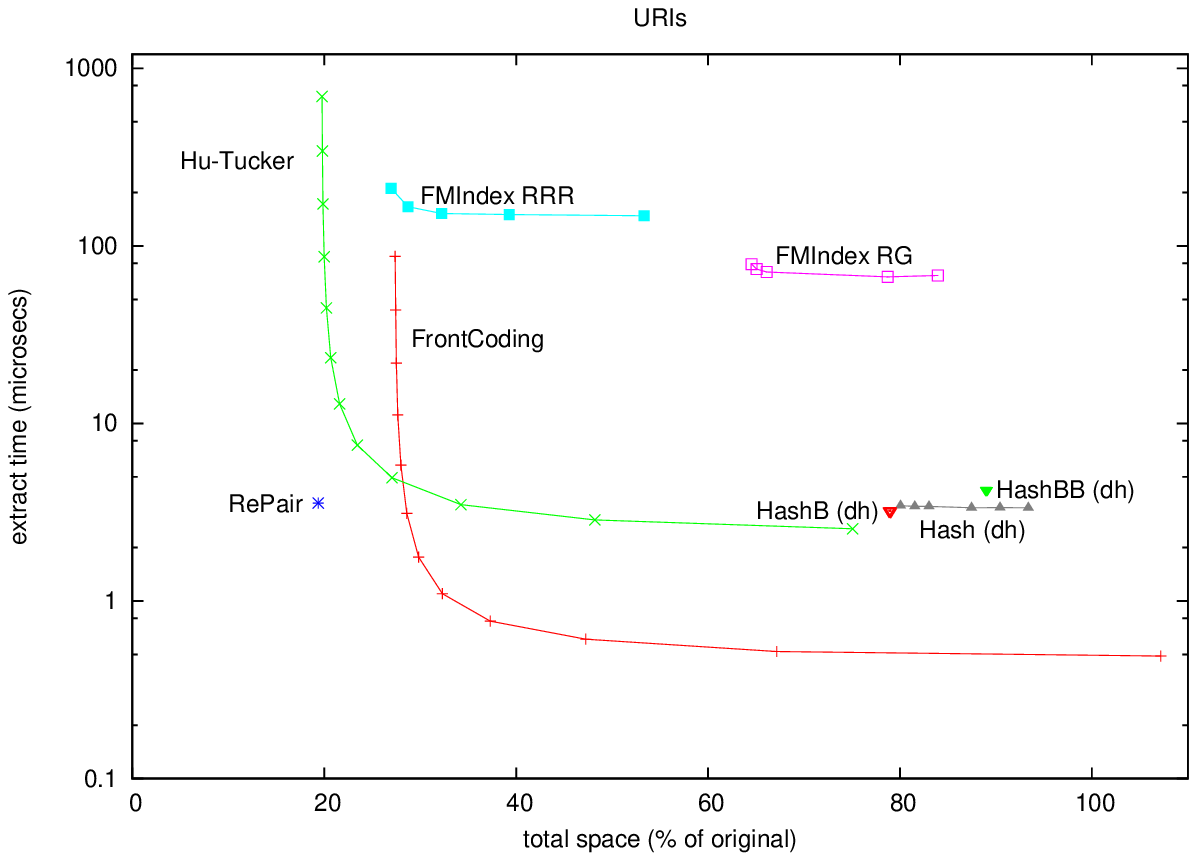}
  \smallskip
\label{fig:exp}
\caption{Locate times (left) and extract times (right) for the different
methods as a function of their space consumption.}
\end{figure}

Figure~\ref{fig:exp} shows our results. 
Most methods are drawn as a line that corresponds to their main space/time
tuning parameter. On the left we show locate time for successful searches; 
the plots for unsuccessful searches are very similar and omitted for lack of
space. On the right we show extraction times. Time is shown in microseconds
and space as a percentage of the space required by concatenating the
plain strings. Since, despite the advantages of linear probing in this
scenario, double hashing was always better, we only plot the latter.

{\em Front-Coding} with Hu-Tucker compression shows to be an excellent choice in 
all cases, achieving good time performance and the least space usage (only
beaten by {\em Re-Pair} on URLs). The folklore {\em Front-Coding}, without compression, 
is almost everywhere dominated by the compressed variant. {\em Re-Pair} achieves
the least space on URLs, yet it is significantly slower than compressed 
{\em Front-Coding}. On the shorter-string dictionaries (Words and DNA), {\em
Re-Pair} does
not compress well. {\em HashBB} performs better in space than {\em HashB} when the
strings are short, otherwise the bitmap becomes too long. It is never,
however, clearly the best alternative. {\em HashB} and {\em Hash} excell in time with
short strings when much space is used (nearly 100\%), yet {\em HashB} is never 
much better than {\em Hash}.

For extracting, the map is dominated by {\em Front-Coding}, in plain or
compressed form (the plain folklore variant is more relevant in this case).
Still {\em Re-Pair} achieves minimum space on URLs.

Another loser in this comparison is the {\em FM-index}. It supports, however, 
more complex searches than the basic ones we have considered.
We discuss this next.

\section{Final Remarks}

Prefix search, that is, finding the dictionary strings that start with a given
pattern, is easily supported by the methods we have explored, except hashing.
Other variants that can likewise be supported are of interest for Internet
routing tables: find the dictionary string that is the longest prefix of the
pattern.

Other searches of interest are only supported by the FM-index:
Find the dictionary strings that contain a
substring, or that have a given prefix and a given suffix \cite{FV10}.
The FM-index also supports approximate searches \cite{RNOM09}. Most other
approximate matching indexes require much extra space \cite{Sun08}.

Several optimizations to our methods are possible, for example using
a compressed rule representation for Re-Pair \cite{GN07}. We have also not 
explored adding compressed tries \cite{ACNS10} to speed up the binary searches
of Front Coding or Re-Pair. Also, compressed
suffix trees \cite{CNsea10} on top of the FM-index could speed it up.

\bibliographystyle{plain}
\bibliography{paper}

\end{document}